\documentclass[11pt,twoside]{article}
\usepackage{asp2006}
\usepackage{epsfig}
\usepackage{epsf}
\usepackage{graphics}
\usepackage{lscape}
\begin{document}
\title{Shear Flows Driven by the Lorentz Force:
An Energy Source for Coronal Mass Ejections and Flares}

\author{W. Manchester IV,\altaffilmark{1}}

\altaffiltext{1}{Center for Space Environment Modeling, University of Michigan, 2455 Hayward Drive, Ann Arbor, MI 48109, USA}

\begin{abstract}
Shear flows have been prescribed in numerical models of coronal mass ejections 
and flares for decades as a way of energizing magnetic fields to erupt.
While such shear flows have long been observed in the solar atmosphere,
until recently, there was no compelling physical explanation for them.
This paper will discuss the discovery that such shear flows are
readily explained as a response to the Lorentz force that naturally
occurs as bipolar magnetic fields emerge and expand in a
gravitationally stratified atmosphere.  It will be shown that shearing
motions transport axial flux, and magnetic energy from the
submerged portion of the field to the expanding portion, strongly
coupling the solar interior to the corona.  This physical process
explains active region shear flows and why the magnetic field is
found to be nearly parallel to photospheric polarity inversion
lines where prominences form.  Finally, shear flows driven by the
Lorentz force are shown to produce a loss of equilibrium and eruption
in magnetic arcades and flux ropes offering a convincing explanation
for CMEs and flares.
\end{abstract}

\section{Introduction}
Coronal mass ejections (CMEs) are energetic expulsions
of plasma from the solar corona that are driven by the
release of magnetic energy typically in the range of $10^{32-33}$ ergs.
The majority of CMEs originate from the eruption of pre-existing
large-scale helmet streamers \citep[]{WM_hundhausen93}.
Less common fast CMEs typically come from smaller, more
concentrated locations of magnetic flux referred to as
active regions.  In this case, the CMEs often occur shortly
after the flux has emerged at the photosphere, but can
also happen even as the active region is decaying.
While CMEs occur in a wide range of circumstances,  
all CMEs originate above photospheric magnetic
polarity inversion lines (neutral lines), which exhibit strong
magnetic shear.  Shear implies that the magnetic field has a strong
component parallel to the photospheric line that separates magnetic
flux of opposite sign, and in this configuration, the field
possesses significant free energy.  In contrast, a potential field
runs perpendicular to the inversion line and has no free energy.

There is enormous evidence for the existence of highly sheared
magnetic fields associated with CMEs and large flares.
At the photosphere, magnetic shear is measured directly
with vector magnetographs \citep[e.g.][]{WM_hagyard84, WM_zirin93,
WM_falconer2002, WM_yang2004, WM_liu2005}.
Higher in the atmosphere, the magnetic field is difficult to
measure directly, but its direction may be inferred from
plasma structures formed within the field.
Seen in chromospheric H$\alpha$ absorption, filaments form only
over photospheric inversion lines \citep[]{WM_zirin83} along which 
the magnetic field is nearly parallel \citep[]{WM_leroy89}.
Fibrils and H$\alpha$ loops that overlay photospheric bipolar
active regions are also indicative of magnetic shear \citep[]{WM_foukal71}.
Comparisons between vector magnetograms and H$\alpha$ images
show that the direction of the sheared photospheric magnetic field
coincides with the orientation of such fibril structures
\citep[]{WM_zhang95}.  Higher in the corona, evidence of magnetic
shear is found in loops visible in the extreme ultraviolet
\citep[]{WM_liu2005} and X-ray sigmoids \citep[]{WM_moore2001}.  These
structures run nearly parallel to the photospheric
magnetic inversion line prior to CMEs, and are followed by
the reformation of closed bright loops that are much more
potential in structure.  Finally, observations 
by the Transition Region and Coronal Explorer (TRACE)
show that 86 percent of two-ribbon flares show a strong-to-weak
shear change of the ribbon footpoints that indicates the eruption
of a sheared core of flux \citep{WM_su2007}.

Sheared magnetic fields are at the epicenter of solar eruptive behavior.
Large flares are preferentially found to occur along the most highly
sheared portions of magnetic inversion lines \citep[]{WM_hagyard84, WM_zhang95}.
More recent analysis by \citet{WM_schrijver2005} found that shear
flows associated with flux emergence drove enhanced flaring.
Similarly, active region CME productivity is also strongly
correlated with magnetic shear as shown by
\citet{WM_falconer2001, WM_falconer2002, WM_falconer2006}.
It is not coincidental that large flares and CMEs are strongly
associated with filaments, which are known to form only along
sheared magnetic inversion lines \citep[]{WM_zirin83}.
The buildup of magnetic shear is essential for
energetic eruptions, and for this reason, it is of fundamental
importance to understanding solar activity.

Currently the majority of CME initiation models rely on the
application of artificially imposed shear flows.  Examples of
such models include 
\citet{WM_antiochos99, WM_mikic94, WM_guo98, WM_amari2003}. 
Until recently, there was no theory  
to account for these large scale shear flows.  
In this paper, we discuss a series of simulations that illustrate a 
physical process by which these shear flows are self-organized
in emerging magnetic fields.  We will give a close comparison of these
simulations in the context of new observations that make a more complete
and compelling picture of a fundamental cause of eruptive solar 
magnetic activity.

\section{Simulations}
\begin{figure*}[ht!]
\begin{center}
\includegraphics[angle=0,scale=0.85]{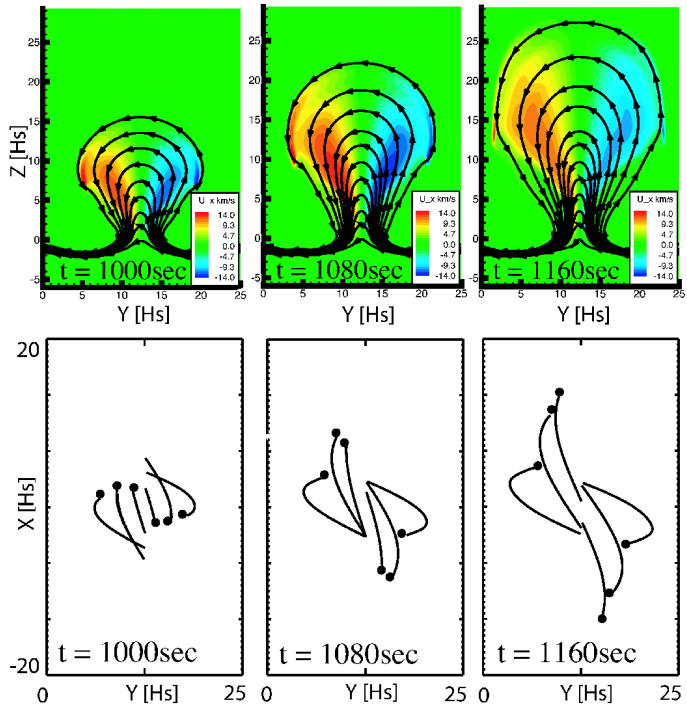}
\end{center}
\caption{Magnetic flux emerging from a horizontal layer into the 
corona in the form of an axisymmetric arcade. 
The top row illustrates the time evolution of the horizontal shear
velocity, $U_x$, shown in color with field lines confined to the
plane drawn black.  The bottom row shows the evolution of 
fully three-dimensional magnetic field lines viewed from above projected 
onto the horizontal $x-y$ plane.  The footpoints of the field lines at
the photosphere are shown as black dots.  The shear flow is 
clearly manifest in the footpoint motion.  The magnetic field near the
center of the arcade evolves to be nearly parallel to the inversion
line, while field lines near the periphery of the system are more
nearly orthogonal.}
\label{layer}
\end{figure*}

Shearing motions and magnetic field alignment with the polarity inversion 
line, so frequently observed in active regions, are readily explained
as a response to the Lorentz force that arises when magnetic flux 
emerges in a gravitationally stratified atmosphere.   The motions 
take the form of large-amplitude shear Alfv\'en waves in which the
magnetic tension force drives horizontal flows in opposite directions
across the polarity inversion line.  The physical process was first shown by
\citet{WM_manchester2000a, WM_manchester2000b, WM_manchester2001} and found 
in later simulations by \citet{WM_fan2001, WM_magara2003, WM_archontis2004}.
As will be discussed in greater detail, what ultimately produces the 
shearing Lorentz force is the nonuniform expansion of the emerging 
magnetic field.

The shearing process was first simulated by \citet{WM_manchester2001}, 
with a two-and-a-half-dimensional (2.5D) simulation of flux emergence 
from a horizontal magnetic layer placed 2-3 pressure scale heights 
below the photosphere.  The crucial aspect of this simulation is that the 
magnetic field is initially in a {\it sheared} configuration oriented at a 45
degree angle to the plane of variation.  In this case, for spatial 
and temporal variations of the instability described by 
$e^{i(\bf{k} \cdot r - \omega t)}$,  $\bf{k}$ is oblique to
the magnetic field ($\bf{B}$).  This distinguishes the mode of 
instability from a Parker mode \citep[]{WM_parker66} in which
$\bf{k}$ is parallel to $\bf{B}$ or an interchange mode in which $\bf{k}$
is perpendicular to $\bf{B}$.  With $\bf{k}$ oblique 
to $\bf{B}$, the instability can be thought of as a mixed mode,
\citep[e.g.][]{WM_cat90, WM_matsumoto93, WM_kusano98}.  In this 
case, there will be a component of the Lorentz force out of 
the plane of variation any where the magnetic component out of the plane
($B_x$ in the chosen coordinates) is not constant along field lines.  
In the simulation by \citet{WM_manchester2001}, it was shown
that the departure of $B_x$ from constant values along field 
lines produced strong shear flows in the emerging flux. 
 
When the magnetic flux rises forming an arcade as shown in Figure 
\ref{layer}, expansion causes the magnitude of $B_x$ to decrease in
the arcade.  The gradient in $B_x$ along field lines results
in the Lorentz force that drives the shear flows.
Examining the expression for the $x$ component of the Lorentz
(tension) force, $F_{x} = \frac{1}{4 \pi} \nabla B_{x} \cdot ({\bf B_y + B_z})$,
we see the reason for the shear.  The gradient in $B_x$ is
negative moving up the arcade in the direction of ${\bf B}$ on
one side of the arcade, while on the opposite side, the gradient
of $B_x$ along ${\bf B}$ is positive. 

\begin{figure*}[ht!]
\begin{center}
\includegraphics[angle=0,scale=0.60]{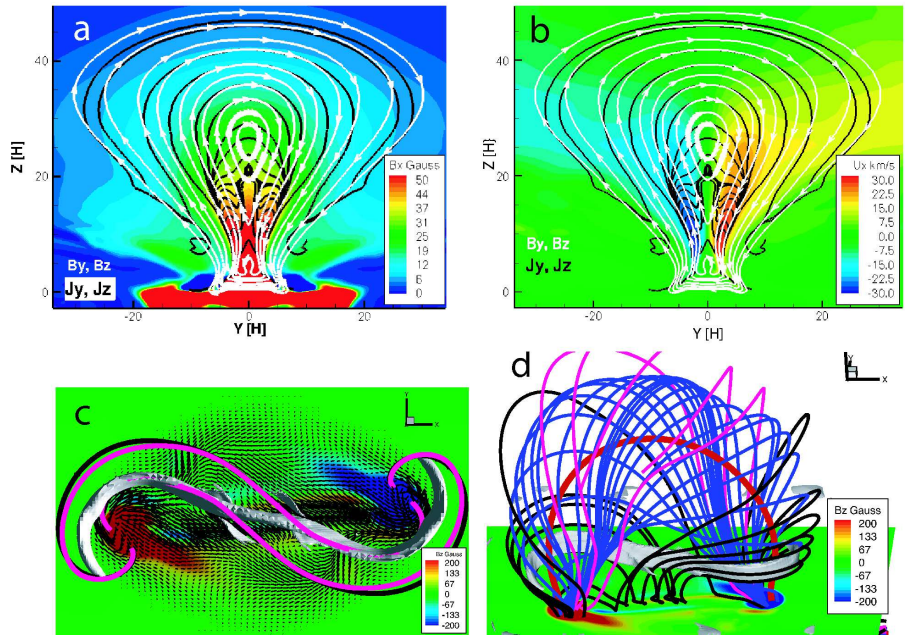}
\end{center}
\caption{
Partial eruption of a three-dimensional emerging magnetic flux rope.
Panel (a) illustrates the Lorentz force with magnetic stream lines
(confined to the $y-z$ plane at the central cross section of the rope)
shown with white lines while black lines show the direction of the
current density.  The existence of the Lorentz force $(\bf{j \times B})$
out of the plane is clearly seen
where the field and current density are oblique.
The magnetic field crosses $\bf{j}$ in opposite directions on
opposite sides of the rope producing the shear flow.
The large vertical gradient in the axial component
(shown in color) produces the horizontal cross-field current.
Panel (b) shows the horizontal shear velocity, which clearly occurs
where $\bf{j}$ and $\bf{B}$ are non-parallel.
The bottom panels show the resulting eruption of the flux rope.
At the photosphere, the vertical magnetic field strength is shown in color
and the horizontal direction is shown with vectors in Panel (c).
Here, the current sheet (where the magnetic field is reconnecting) is shown
with a gray isosurface, along with magnetic field lines entering the
sheet.  Coronal field lines and the current sheet
form a sigmoid structure that runs nearly parallel to the highly sheared
photospheric magnetic field.  Panel (d) shows the flux rope splitting
apart at the current sheet.  The upper part of the rope erupts into the
corona (blue and red lines) while the lower part forms deep dips that
remain just above the photosphere filled with dense plasma.}
\label{rope}
\end{figure*}

These shear flows are illustrated in color in the top row 
of Figure \ref{layer}.  Here, we see the persistence of shear flows
as the magnetic field (drawn with black lines) expands in the 
atmosphere.  The magnitude of the shear flow velocity reaches a peak of 15 km/s
high in the arcade, 3-4 km/s at the photosphere and approximately
1 km/s below the photosphere.  The maximum shear speed typically
reaches half the value of the local Alfv\'en speed.  The bottom
row of Figure \ref{layer} shows the time evolution of magnetic field lines
integrated in three-dimensional space, and shown from above
projected onto the $x-y$ plane.  Footpoints of the field lines
are shown with black dots.  These panels show the shear displacement 
of the footpoints and how the magnetic field evolves to be parallel 
to the polarity inversion line.  Note that the 
field is most highly sheared in close proximity to the polarity
inversion line, and grows more nearly perpendicular to the line
with greater distance from it. 

This same shearing process is found in more complex simulations such as
three-dimensional emerging flux ropes \citep[]{WM_fan2001, WM_magara2003,
WM_manchester2004, WM_archontis2004}.  Panel (a) of Figure \ref{rope} shows  
in greater detail than before the structure of the magnetic field and current
systems in the emerging flux rope of \citet{WM_manchester2004} that leads
to the Lorentz force, which drives the shear flows.  Here, the axial field 
strength of the rope is shown in color, while the direction of the magnetic
field and current density (confined to the plane) are shown with white
and black lines respectively on the vertical plane that is at the central cross
section of the flux rope.  The large expansion of the 
upper part of the flux rope produces a vertical gradient in the 
axial component of the magnetic field $(B_x)$ that results in a horizontal 
current $(j_y)$.  The magnetic field crosses the current in opposite
direction on opposite sides of the flux rope.  This produces the 
Lorentz force out of the plane that reverses direction across the 
flux rope, and drives the horizontal shear flow.  Panel (b) shows
the magnitude of these shear flows on the same central cross section 
of the flux rope.  The shear flow is greatest precisely where the
field is most nearly perpendicular to the current reaching a magnitude
of approximately 30 km/s.  At this time, the flux rope has risen twice
as high as the flux that emerged from the magnetic layer and similarly
the shear flow in the rope is also twice as fast.  

\begin{figure*}[ht!]\begin{center}\includegraphics[angle=0,scale=0.80]{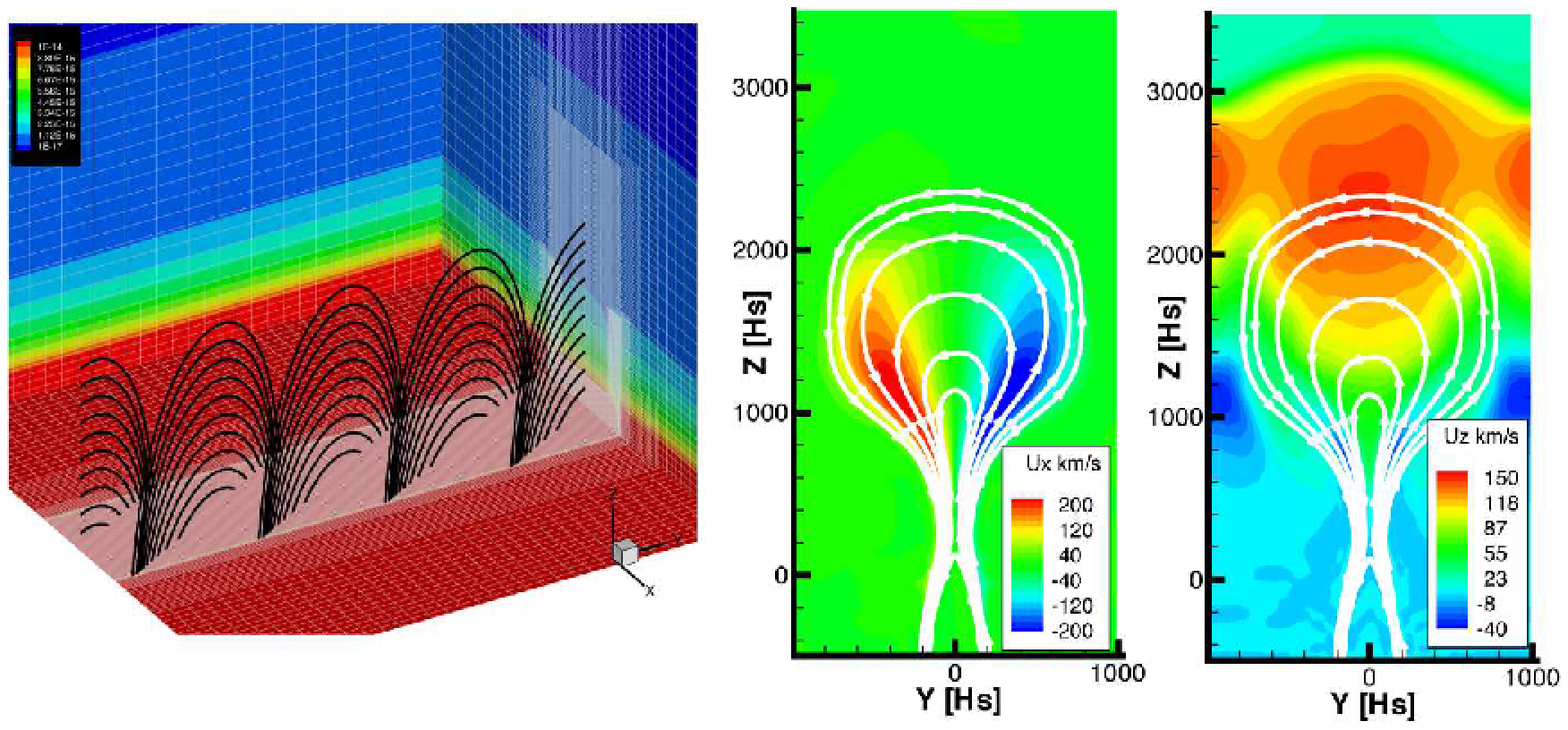}
\end{center}
\caption{Results of the magnetic arcade simulation.  The initial state
is shown on the left with the three-dimensional arcade field lines 
drawn in black, and the density and numerical grid are shown on the outer
boundary of the computational domain.  The middle and right panels,
respectively, show the shear and vertical velocities in color at the
central vertical plane of the simulation.  Field lines (confined
to the plane) are drawn white. Note the much larger scale of the
arcade compared to the flux emergence simulations along with the much
higher velocities.} 
\label{arcade}
\end{figure*}

In the case of \citet{WM_manchester2004} the flux rope erupted, which
was not found in earlier simulations such as \citep[]{WM_fan2001}.  This
eruption resulted from high speed shear flows close to the neutral
line that formed a highly sheared core that lifted off, and was
followed by magnetic reconnection.  The bottom panels show the magnetic
and current structure at the time of the eruption.  Panel (c) gives a
view of the system that shows the vertical magnetic field
in color and the horizontal field with arrows at the photosphere.  A 
current sheet is illustrated with an isosurface of current density.
The sigmoid shape of the sheet reflects the shape of both the coronal 
field lines passing close to the sheet as well as the highly sheared
field at the photosphere.  This model strongly suggests that X-ray sigmoids
seen prior to and during CMEs are indicative of a highly sheared magnetic
field geometry.  Panel (d) shows that at this current sheet,  
reconnection is taking place that separates the upper part of the 
rope that is erupting from the lower part that has V-shaped field lines
full of dense plasma that remains attached to the photosphere. 
Such flux separations in CMEs have been suggested by \citet{WM_gilbert2000}
based on observations of filament eruptions.

Finally we discuss a numerical simulation of an arcade eruption.  This
simulation is essentially identical to the simulation of 
\citet{WM_manchester2003}, with the exception that it is carried out 
in fully three-dimensional space with the BATS-R-US MHD code developed
at the University of Michigan.  The left panel of Figure \ref{arcade} 
shows the initial state of the simulation with field lines drawn in 
black showing the sheared magnetic arcade.  On the boundaries of the
computational domain, the plasma mass density is shown in color along
with the computational grid shown with white lines.  The center and 
right panels show the shear and vertical velocities respectively in
color on the central plane of the simulation.  The magnetic field 
confined to the plane (ignoring the component out of the plane) is 
drawn with white lines in both panels.  In these panels, we find that
the arcade erupts very violently with a shear flow that reach 
a magnitude of 200 km/s.  The arcade rises at a peak velocity of 
150 km/s, and drives a shock ahead of it in the corona. 

In this simulation, the eruption is caused by shear flows driven by
the Lorentz force.  A shearing catastrophe occurs when the $B_x$ 
component of the magnetic field can not be equilibrated along field lines.
This simulation is different  
from the pervious two discussed in that it does not treat flux emergence
through the photosphere, but only models the coronal plasma at a temperature
of one million degrees.  Not needing to resolve the photospheric pressure
scale hight allows much larger cells and a computational domain that is
25 times larger than that of the previous two simulations.  The result 
of this coronal arcade simulation is an eruption that is 5 times faster
than the flux rope eruption, and extends out to half a solar radius 
above the surface.  This progression of eruption 
velocity with the size of the flux system being treating offers
compelling evidence that shear flows driven by the Lorentz are
capable of producing fast CMEs from large active regions. 

\section{Discussion}
The build up of magnetic energy in active regions is essential
to the onset of CMEs and flares.  The magnetic stress must pass
from the convection zone into the corona in the form of non-potential
fields, and effectively couple layers of the atmosphere.
Observations show that such non-potential fields occurs
along magnetic polarity inversion lines where the magnetic
field is highly sheared.  The evidence that magnetic shear is
essential to CMEs and flares is provided by the very strong correlation
between photospheric shear flows, flux emergence and the onset of CMEs and
large flares \citep[]{WM_meunier2003, WM_yang2004, WM_schrijver2005}.  These 
shear flows are found to be strongest along the magnetic inversion line
precisely where flares are found \citep[]{WM_yang2004, WM_deng2006}.
This complements earlier evidence that in the case of two-ribbon flares
\citep[]{WM_zirin84}, prior to the eruption, H$\alpha$ arches
over the inversion line are highly sheared, afterward, the
arches are nearly perpendicular to the inversion line.
\citet{WM_su2006} found a similar pattern of magnetic shear loss
in the apparent motion of footpoints in two-ribbon flares.
There are even now observations of subphotospheric shear flows
with a magnitude of 1-2 km/s 4-6 Mm below the photosphere that
occur during flux emergence \citep[]{WM_kosovichev2006}, which 
simulations show is fully consistent with the Lorentz force driving 
mechanism \citep[]{WM_manchester2007}. 

These ubiquitous shear flows and sheared magnetic fields so strongly
associated with CMEs are readily explained by the Lorentz force
that occurs when flux emerges and expands in a gravitationally 
stratified atmosphere.  This physical process explains and synthesizes
many observations of active regions and gives them meaning in a larger 
context.  This shearing mechanism explains (1) the coincidence of the 
magnetic neutral line with the velocity neutral line, (2) the impulsive
nature of shearing in newly emerged flux, (3) the magnitude of the 
shear velocity in different layers of the atmosphere, (4) the large
scale pattern of magnetic shear in active regions, (5) the 
transport of magnetic flux, and energy from the convection
zone into the corona, (6) eruptions such as CMEs and flares.
With so much explained, it still remains a numerical challenge to 
model an active region with sufficient resolution to produce a large
scale CME by this shearing mechanism.  The rope emergence simulation
discussed here only produces a flux concentration that is one tenth
the size of an active region, which at this scale simply can not produce 
an eruption the size of a CME.  However, this simulation illustrates the 
basic process by which CMEs and flares 
must occur, and current simulations already show a very
favorable scaling of eruption size.  With increases in computer power,
simulations of flux emergence should soon be producing CME size eruptions
by shear flows driven by the Lorentz force. 

\acknowledgements{
Ward Manchester is supported in part at the University of Michigan 
by NASA SR\&T grant NNG06GD62G.  The simulations shown here 
were performed on NCAR and NASA super computers.}

\end{document}